\documentclass[twocolumn]{aastex63}
\usepackage{graphicx}
\usepackage{booktabs}

\published{October 8, 2020}

\submitjournal{ApJL}

\shorttitle{HAT-P-41b's Atmosphere Revealed}
\shortauthors{Lewis et al.}

\begin{document}

\title{Into the UV: The Atmosphere of the Hot Jupiter HAT-P-41b Revealed}

\correspondingauthor{Nikole K. Lewis}
\email{nikole.lewis@cornell.edu}

\author[0000-0002-8507-1304]{N.~K. Lewis}
\affiliation{Department of Astronomy and Carl Sagan Institute, 
Cornell University, 122 Sciences Drive, Ithaca, NY 14853, USA}

\author[0000-0003-4328-3867]{H.~R. Wakeford}
\affil{School of Physics, University of Bristol, HH Wills Physics Laboratory, Tyndall Avenue, Bristol BS8 1TL, UK}

\author[0000-0003-4816-3469]{R.~J. MacDonald}
\affil{Department of Astronomy and Carl Sagan Institute, Cornell University, 122 Sciences Drive, Ithaca, NY 14853, USA}

\author[0000-0002-8515-7204]{J.~M. Goyal}
\affiliation{Department of Astronomy and Carl Sagan Institute, 
Cornell University, 122 Sciences Drive, Ithaca, NY 14853, USA}

\author[0000-0001-6050-7645]{D.~K. Sing}
\affil{Department of Physics and Astronomy, Johns Hopkins University, Baltimore, MD 21218, USA}
\affil{Department of Earth and Planetary Sciences, Johns Hopkins University, Baltimore, MD 21218, USA}

\author[0000-0003-3726-5419]{J. Barstow}
\affiliation{School of Physical Sciences, Open University, Walton Hall, Kents Hill, Milton Keynes, MK7 6AA UK}
\affiliation{Department of Physics and Astronomy, University College London, Gower St, London WC1E 6BT, UK}

\author[0000-0002-4250-0957]{D. Powell}
\affil{Department of Astronomy and Astrophysics, University of California, Santa Cruz, CA 95064, USA}

\author[0000-0003-3759-9080]{T. Kataria}
\affil{Jet Propulsion Laboratory, California Institute of Technology, 4800 Oak Grove Drive, Pasadena, CA 91109, USA}

\author[0000-0001-6092-7674]{I. Mishra}
\affiliation{Department of Astronomy and Carl Sagan Institute, Cornell University, 122 Sciences Drive, Ithaca, NY 14853, USA}

\author[0000-0002-5251-2943]{M.~S. Marley}
\affil{NASA Ames Research Center, MS 245-3, Mountain View, CA 94035, USA}

\author[0000-0003-1240-6844]{N.~E. Batalha}
\affil{NASA Ames Research Center, MS 245-3, Mountain View, CA 94035, USA}

\author[0000-0002-8837-0035]{J.~I. Moses}
\affil{Space Science Institute, 4765 Walnut Street, Suite B, Boulder, CO 80301, USA}

\author[0000-0002-8518-9601]{P. Gao}
\affiliation{Department of Astronomy, University of California, Berkeley, CA 94720, USA}

\author[0000-0001-6352-9735]{T.~J. Wilson}
\affil{University of Exeter, Physics Building, Stocker Road, Exeter, Devon, Ex4 4QL, UK}

\author[0000-0002-4552-4559]{K.~L. Chubb}
\affiliation{SRON Netherlands Institute for Space Research, Sorbonnelaan 2, 3584 CA, Utrecht, Netherlands}

\author[0000-0001-5442-1300]{T.~Mikal-Evans}
\affil{Kavli Institute for Astrophysics and Space Research, Massachusetts Institute of Technology, 77 Massachusetts Avenue, 37-241, Cambridge, MA 02139, USA}

\author[0000-0002-6500-3574]{N. Nikolov}
\affil{Space Telescope Science Institute, 3700 San Martin Drive, Baltimore, MD 21218, USA}
\affil{Department of Physics \& Astronomy, Johns Hopkins University, Baltimore, MD 21218, USA}

\author{N. Pirzkal}
\affil{Space Telescope Science Institute 3700 San Martin Drive Baltimore, MD 21218, USA} 

\author{J.~J. Spake}
\affil{Division of Geological and Planetary Sciences, California Institute of Technology, Pasadena, CA 91125, USA}

\author[0000-0002-7352-7941]{K.~B. Stevenson}
\affiliation{Johns Hopkins APL, 11100 Johns Hopkins Rd, Laurel, MD 20723, USA}

\author[0000-0003-3305-6281]{J. Valenti}
\affil{Space Telescope Science Institute, 3700 San Martin Drive, Baltimore, MD 21218, USA}

\author{X. ~Zhang}
\affiliation{Department of Earth and Planetary Sciences, University of California, Santa Cruz, CA 95064, USA}

\begin{abstract}

For solar-system objects, ultraviolet spectroscopy has been critical in identifying sources for stratospheric heating and measuring the abundances of a variety of hydrocarbon and sulfur-bearing species, produced via photochemical mechanisms, as well as oxygen and ozone. To date, less than 20 exoplanets have been probed in this critical wavelength range (0.2-0.4~$\mu$m). Here we use data from Hubble’s newly implemented WFC3 UVIS G280 grism to probe the atmosphere of the hot Jupiter HAT-P-41b in the ultraviolet through optical in combination with observations at infrared wavelengths. We analyze and interpret HAT-P-41b’s 0.2-5.0~$\mu$m transmission spectrum using a broad range of methodologies including multiple treatments of data systematics as well as comparisons with atmospheric forward, cloud microphysical, and multiple atmospheric retrieval models. Although some analysis and interpretation methods favor the presence of clouds or potentially a combination of Na, VO, AlO, and CrH to explain the ultraviolet through optical portions of HAT-P-41b’s transmission spectrum, we find that the presence of a significant H$^{-}$ opacity provides the most robust explanation. We obtain a constraint for the abundance of H$^{-}$ , $\log(\rm{H^{-}}) = -8.65 \pm 0.62$ in HAT-P-41b’s atmosphere, which is several orders of magnitude larger than predictions from equilibrium chemistry for a $\sim$1700-1950~K hot Jupiter. We show that a combination of photochemical and collisional processes on hot hydrogen-dominated exoplanets can readily supply the necessary amount of H$^{-}$ and suggest that such processes are at work in HAT-P-41b and many other hot Jupiter atmospheres.

\end{abstract}


\section{Introduction} \label{sec:intro}

For more than 50 years now a variety of space-based observatories have provided a window into the ultraviolet (UV) properties of planetary objects \citep[see review in][]{brosch2006}. The UV provides a unique perspective on a number of physical processes in planetary atmospheres, such as photodissociation and photoionization, as well as containing unique spectral indicators for a broad range of atoms, ions, and molecules. In the solar system the energy source for stratospheric heating is often absorption of solar UV radiation, usually by high altitude molecules or hazes produced by photochemical processes, ozone in Earth's stratosphere being the prototypical example. Other examples include UV absorption by aerosols in the giant planet stratospheres \citep[e.g.][]{zhang2015}. We now know that these UV-driven processes and spectral signatures are not just limited to the solar system. Exoplanets have been observed to also possess stratospheres \citep[e.g.][]{evans2017} and display signatures of a variety of atmospheric aerosols \citep[e.g.][]{sing2016_nature} and ionic species \citep[e.g.][]{sing2019, hoeijmakers2019}. However, the number of exoplanet transits observed at UV {\it and} near-UV (0.2-0.4~$\mu$m) wavelengths is only about a dozen, which limits our ability to fully explore the atmospheric processes shaping these worlds. Here we expand the sample of exoplanets with UV through infrared (IR) observations with our exploration of HAT-P-41b that leverages data using a new observing strategy with the {\it Hubble Space Telescope}.

HAT-P-41b is an inflated `hot Jupiter' (0.8~M$_J$, 1.7~R$_J$, 1940~K) discovered orbiting an F-type star by \citet{hartman2012}. The quiet nature of the host star, highly inflated planetary atmosphere, and short orbital period make HAT-P-41b an ideal target for spectroscopic analysis to probe the physics and chemistry at work in this planet's atmosphere. For this reason, HAT-P-41b was targeted for five observations as part of the Panchromatic Exoplanet Treasury Survey (PanCET GO-14767, PIs: Sing \& Lopez-Morales) using three modes on Space Telescope Imaging Spectrograph (STIS) (E230M, 2$\times$G430L, G750L gratings) and the G141 infrared grism on Wide Field Camera 3 (WFC3), that would provide a spectrum from 0.22--1.7\,$\mu$m. To further leverage \textit{Hubble}'s spectroscopic capabilities, our team also selected HAT-P-41b as the prime target to test the use of WFC3's UVIS G280 grism (GO-15288, PIs: Sing \& Lewis) on exoplanet time series studies. WFC3's UVIS grism provides continuous coverage in a single observation from the UV to the optical and has the potential to replace the equivalent three modes on STIS traditionally used. While the WFC3/UVIS grism presents some challenges for both observational strategies and data reduction, it is able to produce high precision spectroscopy from the UV through visible (0.2--0.8\,$\mu$m) at higher resolution than the combination of the optical STIS modes (see \citealt{wakeford2020} for full details). 

Here we present a detailed exploration of the physics and chemistry that is shaping the transmission spectrum of HAT-P-41b from 0.2-5.0~$\mu$m. We have leveraged data from the newly commissioned WFC3-UVIS G280 mode (0.2-0.8~$\mu$m) and combined with observations 
of the system from WFC3-IR G141 mode (1.1-1.7~$\mu$m) and {\it Spitzer's} Infrared Array Camera (IRAC) (3.6 and 4.5~$\mu$m channels). We present an independent analysis of the PanCET WFC3 G141 data that was previously published as part of a population study by \citet{Tsiaras2018}, and combine it with the data published by \citet{wakeford2020}. In interpreting our transmission spectrum of HAT-P-41b we employ a broad range of analysis tools and techniques including comparisons with atmospheric forward model grids, 
three-dimensional general circulation models, aerosol microphysics models, and three different atmospheric retrieval tools. We apply this bevvy of analysis tools to transmission spectra of HAT-P-41b that utilize two different reduction methods for the WFC3 UVIS G280 data to test the robustness of our interpretation to data reduction method employed. This work highlights that our interpretation of exoplanet atmospheric characterization observations are served by exploring multiple reductions of the same data and multiple analyses that can provide complementary views of the processes shaping exoplanetary atmospheres.

\section{Observations} \label{sec:data}

Observations of the transiting hot Jupiter HAT-P-41b were obtained with \textit{Hubble} and \textit{Spitzer} Space Telescopes over three different observing programs to construct the transmission spectrum from 0.2--5\,$\mu$m. Details of the reduction and data analysis of the Hubble WFC3 UVIS G280 grism spectra (GO-15288, PI: Sing \& Lewis) and Spitzer IRAC 3.6 and 4.5\,$\mu$m (13044, PI: Deming) data can be found in \citet{wakeford2020}. In that work they present different analysis methods on the newly implemented UVIS grism mode, that result in two independent but consistent transmission spectra of approximately 60 measurements in 10\,nm bins from 0.2--0.8\,$\mu$m. The \textit{Spitzer} measurements at 3.6 and 4.5\,$\mu$m are used to further constrain the system parameters, a/R$_*$, inclination, and orbital period which are then fixed in the spectroscopic analysis to prevent arbitrary offsets between the measurements. Here we outline the data analysis performed on the Hubble WFC3 IR G141 transmission spectra taken as part of the \textit{Hubble} PanCET program (GO-14767, PI: Sing \& Lopez-Morales) used to measure the planet's near-IR spectrum, which is critical to the full interpretation of the atmosphere.

A single  transit of HAT-P-41b was observed on 2016 October 16 using \textit{Hubble}'s WFC3 IR G141 grism, 1.1--1.7\,$\mu$m. The observations were conducted in spatial scan mode on the 256 subarray with an exposure time of 81 seconds and a scan rate of 0.072''/s resulting in a scan length of approximately 45 pixels on the detector. The nearby companion is also visible in the scan and overlaps with the target spectrum by approximately 30 pixels. We therefore use the difference imaging technique (e.g., \citealt{kreidberg2014a,evans2016}) to sample the target scan (NSAMP = 12) and reconstruct the target spectrum without the influence of the companion star. To ensure a robust interpretation of the full planetary transmission spectrum we opt not to use the published G141 transmission spectrum presented by \citet{Tsiaras2018} as \citet{wakeford2020} provides updated HAT-P-41b system parameters that should be used consistently across reductions of the {\it Hubble} WFC3/UVIS G280, WFC3/IR G141 and {\it Spitzer} IRAC 3.6 and 4.5\,$\mu$m observations used to construct the UV through IR spectrum considered in this study.

We first analyse the broadband lightcurve from 1.1--1.7\,$\mu$m prior to dividing the spectrum into multiple spectroscopic bins. The observation spanned 5 HST orbits, in this analysis we follow standard practices and discard the first orbit and the first exposure in each orbit as they are subject to additional systematics (e.g., \citealt{deming2013,sing2016_nature}). We analyze the transit time-series data using the instrument systematics marginalization technique outlined by \citet{wakeford2016}, that has been successfully applied to a range of datasets (e.g., \citealt{kilpatrick2018, sing2016_nature, Wakeford2017Science, wakeford2018,wakeford2020}). For a consistent analysis between all datasets we fix the system parameters determined in \citet{wakeford2020}; orbital period\,=\,2.69404861 days, inclination\,=\,89.17$^\mathrm{o}$, and a/R$_*$\,=\,5.55. We also apply the same technique used to calculate the limb-darkening coefficients over the desired wavelengths; 3D stellar models using a 4-parameter non-linear limb-darkening law. This ensures that there are no offsets between each dataset that would compromise the interpretation, which would be the case if using previously published analysis of these data from \citet{Tsiaras2018} that employ different assumptions (see Appendix \ref{appendix_0} for more details).

\begin{figure*}[ht!]
\includegraphics[width=\textwidth]{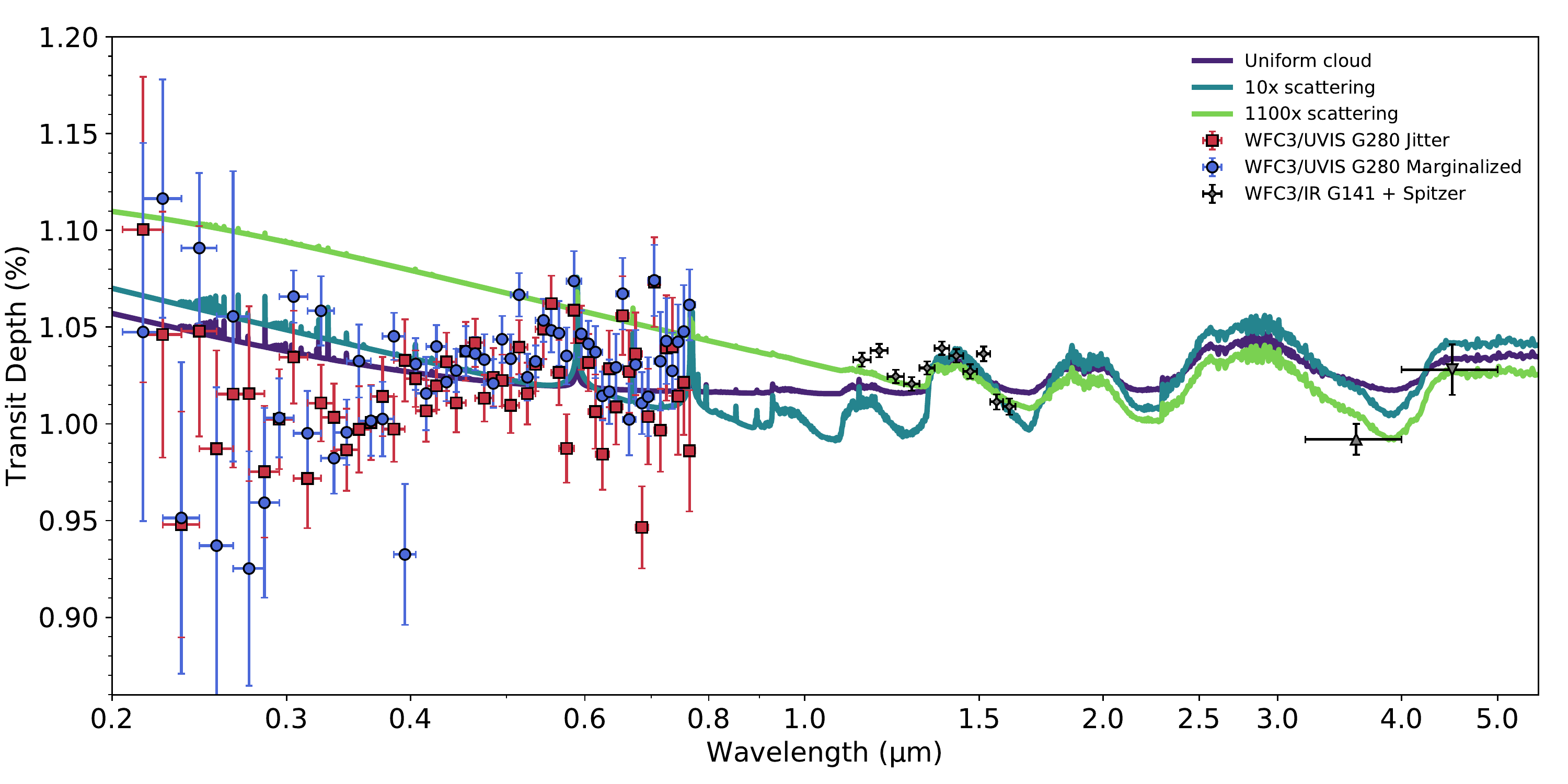}
\caption{HAT-P-41b's transmission spectrum from 0.2--5.0~$\mu$m (red/blue/grey points) obtained via observations with {\it Hubble's} WFC3 UVIS G280 (0.2--0.8~$\mu$m) and IR G141 (1.1--1.7~$\mu$m) grisms and {\it Spitzer's} IRAC Ch1 (3.6~$\mu$m) and Ch2 (4.5~$\mu$m) photometry. We include two reductions of the WFC3 UVIS G280 data that employ marginalization (blue circles) and jitter decorrelation (red squares) treatments of systematics. Theoretical transmission spectra from atmospheric models specific to HAT-P-41b from \citet{goyal2019} are shown for comparison. Atmospheric models with large amounts of 
scattering due to the presence of small particles in the atmosphere (light green line) best match the {\it Hubble} WFC3 IR G141 and {\it Spitzer} Ch1 and Ch2 observations, but fail to match the {\it Hubble} WFC3 UVIS G280 observations where models with low scattering (teal line) and/or a uniform cloud deck (dark purple line) are preferred. This highlights the need for NUV (0.2--0.4~$\mu$m) and optical in additional to IR (1.0-5.0~$\mu$m) observations to robustly probe exoplanet atmospheres.}  
\label{fig:data}
\end{figure*}

We obtained a transit depth precision of 12\,ppm on the broadband lightcurve, with an average precision of 40\,ppm in 47\,nm bins from 1.1--1.7\,$\mu$m. We tested a range of binning options along the spectrum and found the 47\,nm bins to be consistent in the structure of the resultant transmission spectrum while minimizing the scatter of the light curve residuals. Each lightcurve is independently analysed correcting for systematics using a grid of 50 sudo-stochastic polynomial models that account for observatory and instrument based systematics (see \citealt{wakeford2016}). For each lightcurve we calculate the maximum likelihood estimation based on the Akaike information criterion (AIC) for each of the 50 potential systematic models. These are then used as an approximation of the evidence for that systematic correction and converted into a normalized weight. The transit depth measured based on the fit of each systematic model is then marginalized based on the weight assigned to the model such that the marginalized transit depth is representative of the evidence across all 50 models. This method is applied to each spectroscopic light curve measured in each wavelength bin (see spectroscopic light curves in the \href{https://doi.org/10.5281/zenodo.4023155}{supplementary material}). 
The full transmission spectrum we obtain for HAT-P-41b from 0.2--5\,$\mu$m, combining the IR measurements with the UVIS and \textit{Spitzer} data, is presented in Figure~\ref{fig:data}.

\section{Spectral Analysis} \label{sec:analysis}

Our initial comparison of HAT-P-41b's transmission spectrum with theoretical predictions for the planet (Figure~\ref{fig:data}) highlights that further exploration of the physical and chemical processes shaping the spectrum is needed. Here we employ a multiple modelling approach to ensure that we fully explore these processes and our interpretation of HAT-P-41b's spectrum is robust. We first employ one-dimensional (1D) scalable forward models that assume equilibrium chemistry and provide a range of parameterizations to represent the presence of aerosols in the atmosphere. We then leverage predictions from a three-dimensional (3D) general circulation model for HAT-P-41b to guide exploration of aerosol formation using a microphysical model. Finally, we conduct a series of atmospheric retrieval analyses to infer the atmospheric composition of HAT-P-41b. In all our analyses we consider both the marginalization and jitter decorrelation treatments of the systematics for the {\it Hubble} WFC3 UVIS G280 data. Exploration of the similarities and differences in our inferences from these forward and inverse modelling approaches will allow us to robustly identify the physics and chemistry at work in HAT-P-41b's atmosphere. 

\subsection{Comparisons with Forward Models \label{subsec:forward}}

\begin{figure*}[ht!]
\includegraphics[width=\textwidth]{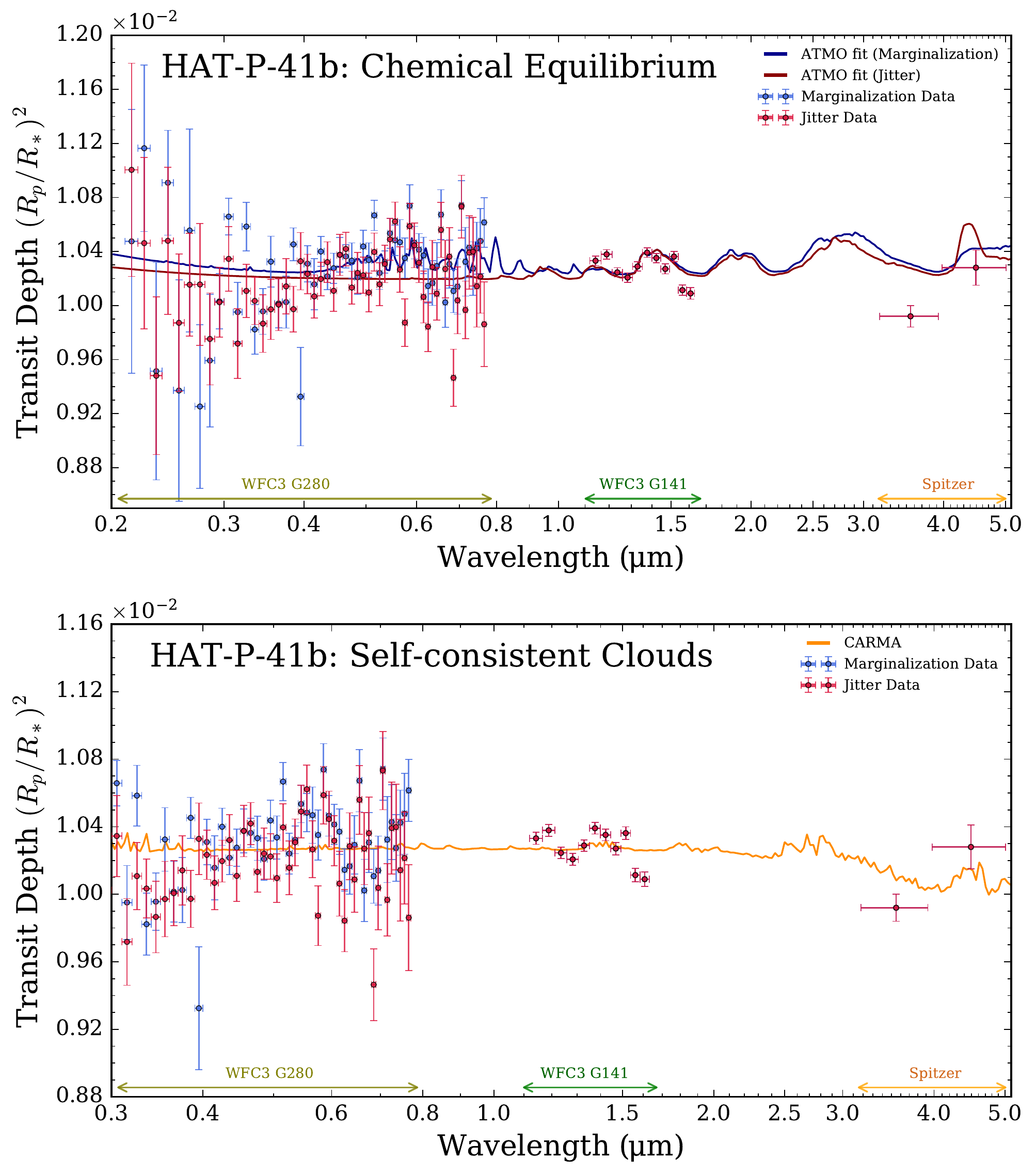}
\caption{Forward model fits to HAT-P-41b's transmission spectrum. Top: chemical equilibrium grid fit. Two independent WFC3 UVIS G280 data reduction techniques are shown: marginalization (blue) and jitter decorrelation (red). For each data reduction, the best fitting model from the \citet{goyal2019} grid is displayed. Both fits favor a cloud deck to explain the continuum UVIS data, while the marginalization fit additionally includes VO opacity to fit optical substructure. Neither model can fit the two reddest WFC3 IR G141 data points nor the 3.6~$\micron$ Spitzer point. Bottom: self-consistent microphysical cloud model fit. Vertical cloud distributions are computed using CARMA \citep{gao2018}, assuming solar metallicity, with transmission spectra computed as in \citet{powell2019}. The assumed temperature and vertical mixing profiles are perturbed from the globally averaged GCM profiles of Figure~\ref{fig:gcm_profiles} to identify the model with minimal residuals. The best-fitting model favors a population of Al$_2$O$_3$ clouds at $\sim 10^{-1}-10^{-3}$~bar. The clouds become optically thin at longer wavelengths, resulting in an improved fit to the 3.6$\micron$ Spitzer point, but struggle to capture the WFC3 G141 H$_2$O feature in the optically thick regime.}
\label{fig:forward_models}
\end{figure*}

We begin our exploration of HAT-P-41b's atmosphere by comparing our observed spectra with a grid of 1D forward models. The \citet{goyal2019} library of `generic' exoplanet transmission spectra spans a broad range of atmospheric compositions, temperature, and aerosol properties, under the assumption of equilibrium chemistry. Our fitting procedure follows a similar methodology to \citet{wakeford2018}, repeated for each WFC3 UVIS G280 data reduction. The resulting best fitting models for the jitter detrending and marginalization reductions are shown in Figure~\ref{fig:forward_models} (top panel). Though the forward model grid can fit most of the observations, mismatches occur in several spectral regions, especially near 1.5 and 3.6~$\micron$. As a result, the $\chi^{2}_{\nu}$ values for this equilibrium model were 2.24 and 2.60 for the jitter and marginalization cases, respectively.

We find slight variations between the preferred forward models for each data reduction. The best fitting model for the marginalization data reduction estimates the atmospheric temperature at the pressures probed via transmission as 1400~K, with a solar metallicity, solar carbon-to-oxygen (C/O) ratio (0.56), and a significant cloud deck. In contrast, the best fitting model for the jitter detrending reduction approach prefers a colder atmosphere (900~K), 10$\times$ solar metallicity, subsolar C/O (0.35), and a significant cloud deck. The fits are broadly consistent within the limits provided by their pseudo-probability distributions\footnote{All our pseudo-probability distributions are available in the online \href{https://doi.org/10.5281/zenodo.4023155}{supplementary material}.}. However, note that the best fitting model in the marginalization case allows for the presence of VO, which can be present under chemical equilibrium assumptions at 1400~K at relevant atmospheric pressures, to explain the spectral structure shortwards of 1~$\mu$m (Figure~\ref{fig:forward_models}, top panel). This illustrates a degree of sensitivity to the chosen data reduction technique - we further quantify this in section~\ref{subsec:retrievals}.

\begin{figure*}
    \centering
    \includegraphics[width=0.49\textwidth]{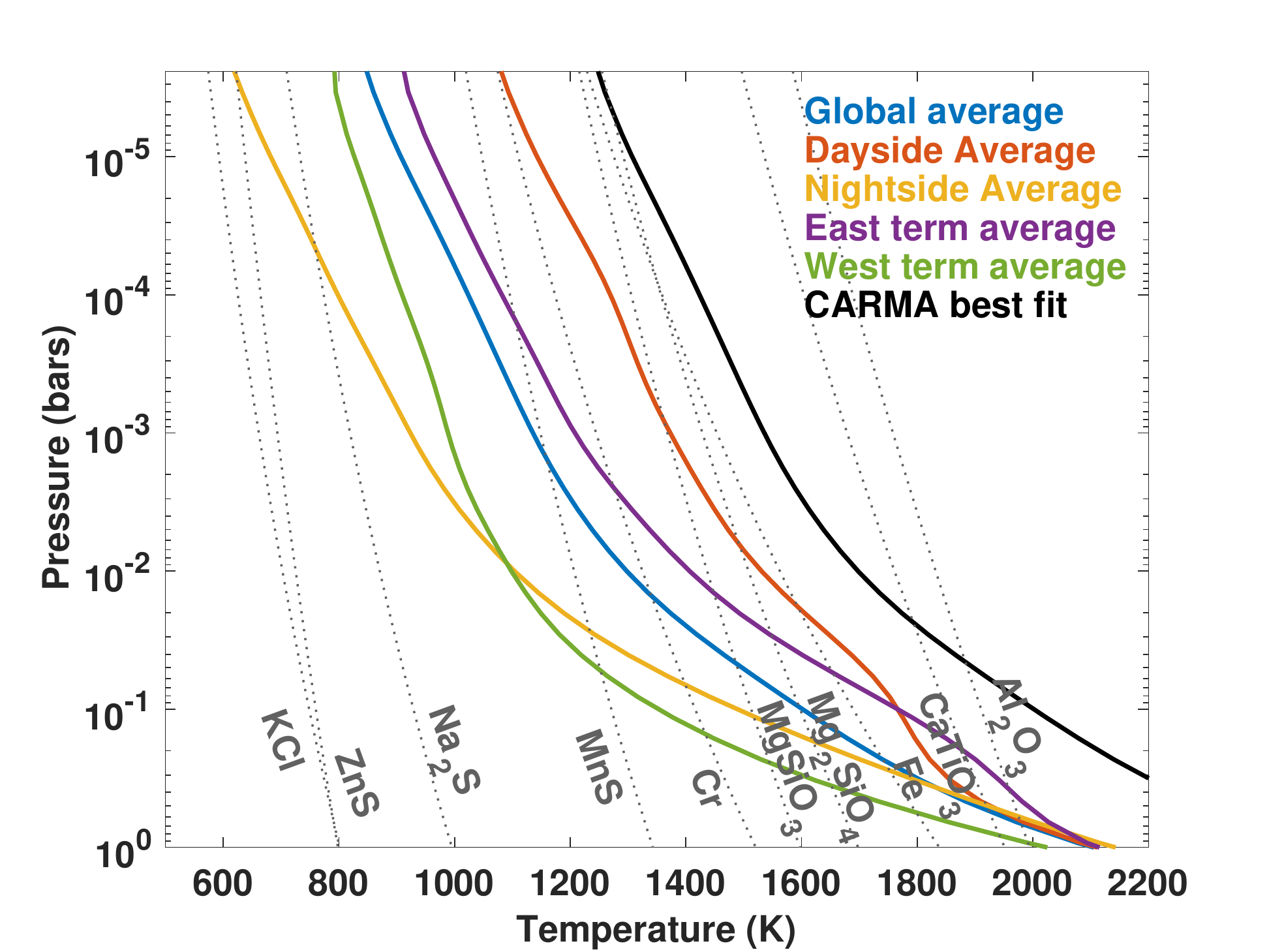}
    \includegraphics[width=0.49\textwidth]{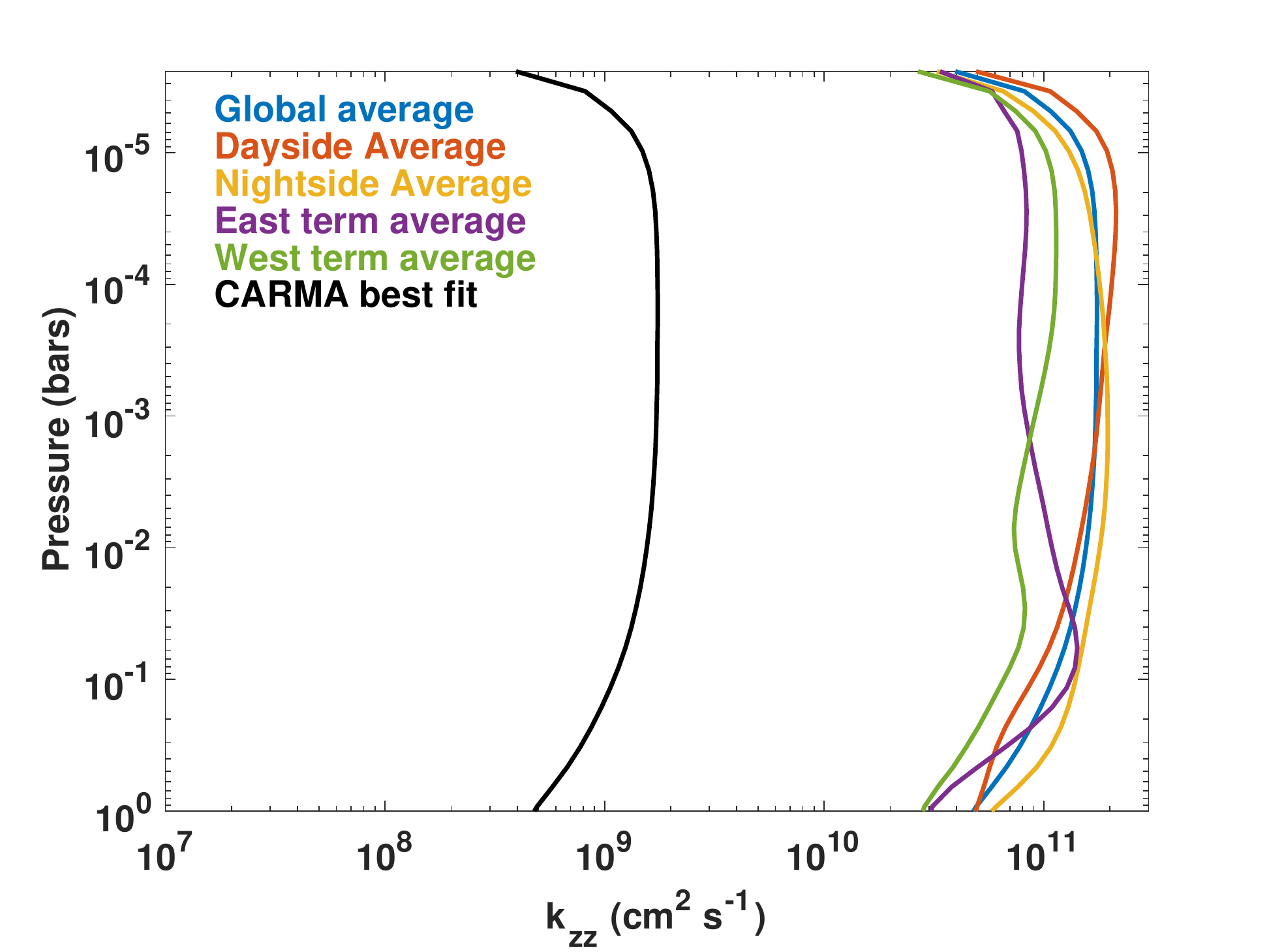}
    \caption{Average pressure-temperature (PT) profiles (left) and vertical diffusion coefficient ($k_{zz}$, right) derived from three-dimensional general circulation model for HAT-P-41b. The average PT profiles intersect 
    with the condensation curves of a number of potential cloud species (dotted lines in left panel), which indicates that clouds could play a critical role throughout HAT-P-41b's atmosphere. The strength of the predicted vertical mixing in HAT-P-41b's atmosphere (right panel) highlights that clouds formed at the bar level or below could be easily advected into
    the observable portion of the atmosphere near a millibar. Profiles for temperature and $k_{zz}$ that provided the CARMA model best fit to the data are shown for comparison.}
    \label{fig:gcm_profiles}
\end{figure*}

Our forward model comparison indicates clouds could play a key role in shaping HAT-P-41b's transmission spectrum. However, our grid includes clouds via a simple cloud deck pressure decoupled from the atmospheric dynamics and thermochemical structure. We therefore verified the plausibility of the formation and advection of clouds in HAT-P-41b's atmosphere by running a general circulation model using the SPARC/MITgcm \citep[e.g.][]{showman2009, kataria2016}. We assume the same physical parameters of HAT-P-41b used throughout this study and a solar metallicity atmospheric composition consistent with the preferred 1D atmospheric models. The resulting globally and spatially-averaged temperature and vertical mixing profiles for the dayside, nightside, and each terminator are shown in Figure~\ref{fig:gcm_profiles}. The temperature profiles of the west (morning; green profiles) and east (evening; purple profiles) terminators cross multiple condensation curves indicating that a broad range of cloud species may be present in HAT-P-41b's observable atmosphere. The $\sim 200$\,K difference between the terminators may drive differing cloud properties on each limb of the planet. With their plausibility and constraints on vertical mixing established, we turn to detailed microphysical modelling to investigate the physical nature and composition of clouds in the atmosphere of HAT-P-41b.

\subsection{Predictions for Cloud Formation \label{subsec:cloud}}

We simulate cloud distributions for HAT-P-41b using the Community Aerosol and Radiation Model for Atmospheres (CARMA). CARMA is a one-dimensional bin-scheme aerosol microphysics model that computes vertical and size distributions of aerosol particles. CARMA solves a discretized continuity equation that accounts for aerosol nucleation, condensation, evaporation, and transport \citep[see][and references therein]{gao2018}. The specific microphysical setup that we use for modeling condensational clouds in this work is described by \citet{powell2019} and has been shown to reproduce trends in Hot Jupiter cloudiness across a broad range of parameter space. In this work we do not vary the microphysical parameters of the condensate species and we assume that the volatile species in the atmosphere have a solar abundance. We post-process these results to calculate transmission spectra using a modified version of Exo-Transmit that includes a Mie theory prescription for the cloud opacities as described by \citet{powell2019}.   

To model the observations of HAT-P-41b, we must choose input planetary properties. As the atmospheric metallicity and microphysical parameters are held constant, the remaining tunable parameters are the temperature profile and the amount of atmospheric vertical mixing. For both parameters, we begin by considering the globally averaged 3D GCM temperature and vertical mixing profile shown in Figure~\ref{fig:gcm_profiles}. We then vary both profiles by a constant factor, as both parameters are not well constrained, to derive cloud properties that give rise to simulated spectra that best match the observations. 

The best-fitting model, shown in Figure~\ref{fig:forward_models} (bottom panel), is 400~K hotter than the globally-averaged temperature profile from the 3D GCM and has two orders-of-magnitude less global mixing. The increase in temperature limits the supersaturation of the condensible species and thus the formation of clouds. This increase in temperature is consistent with temperature variations seen in the GCM, especially limb-to-limb variations \citep[e.g.][]{kataria2016}. Reducing the amount of vertical mixing further limits the formation of clouds as well as the vertical extent in the atmosphere of the cloud particles \citep[see][]{powell2019}. The level of reduction in the vertical mixing calculated from the GCM is consistent with studies of GCM tracer transport where the derived mixing of tracers is commonly two orders of magnitude less efficient than the transport of atmospheric gases \citep[e.g.][]{parmentier2013}. Both of these effects give rise to a population of large aluminum (Al$_2$O$_3$) clouds at $\sim 10^{-1}-10^{-3}$~bar that dominate the cloud opacity. These clouds are optically thick at short wavelengths and optically thin at longer wavelengths \citep{vahidinia2014}. The results of this one-dimensional globally averaged cloud modeling result in a similar spectral fit to the first order forward models considered in section~\ref{subsec:forward} (e.g. the jitter fit in the top panel of Figure~\ref{fig:forward_models}). However, the inclusion of wavelength-dependant cloud opacities improves the fit at 3.6~$\micron$. Nevertheless, this cloud model fails to reproduce the full shape of the H$_2$O feature centered at 1.4~$\micron$. The corresponding $\chi^{2}_{\nu}$ values for this cloud model were 2.67 and 2.85 for the jitter and marginalization cases, respectively. As an additional sanity check, we ran simple Mie theory model comparisons \citep{wakeford2015}, which required large ($\sim$0.5-10~$\mu$m) particles that result in gray opacities out to 2~$\mu$m and miss key molecular absorption features. In seeking a model capable of explaining the observations over the full wavelength range, we turn now to retrieval analyses.


\subsection{Atmospheric Retrieval Analyses \label{subsec:retrievals}}

\begin{figure*}[ht!]
\includegraphics[width=\textwidth]{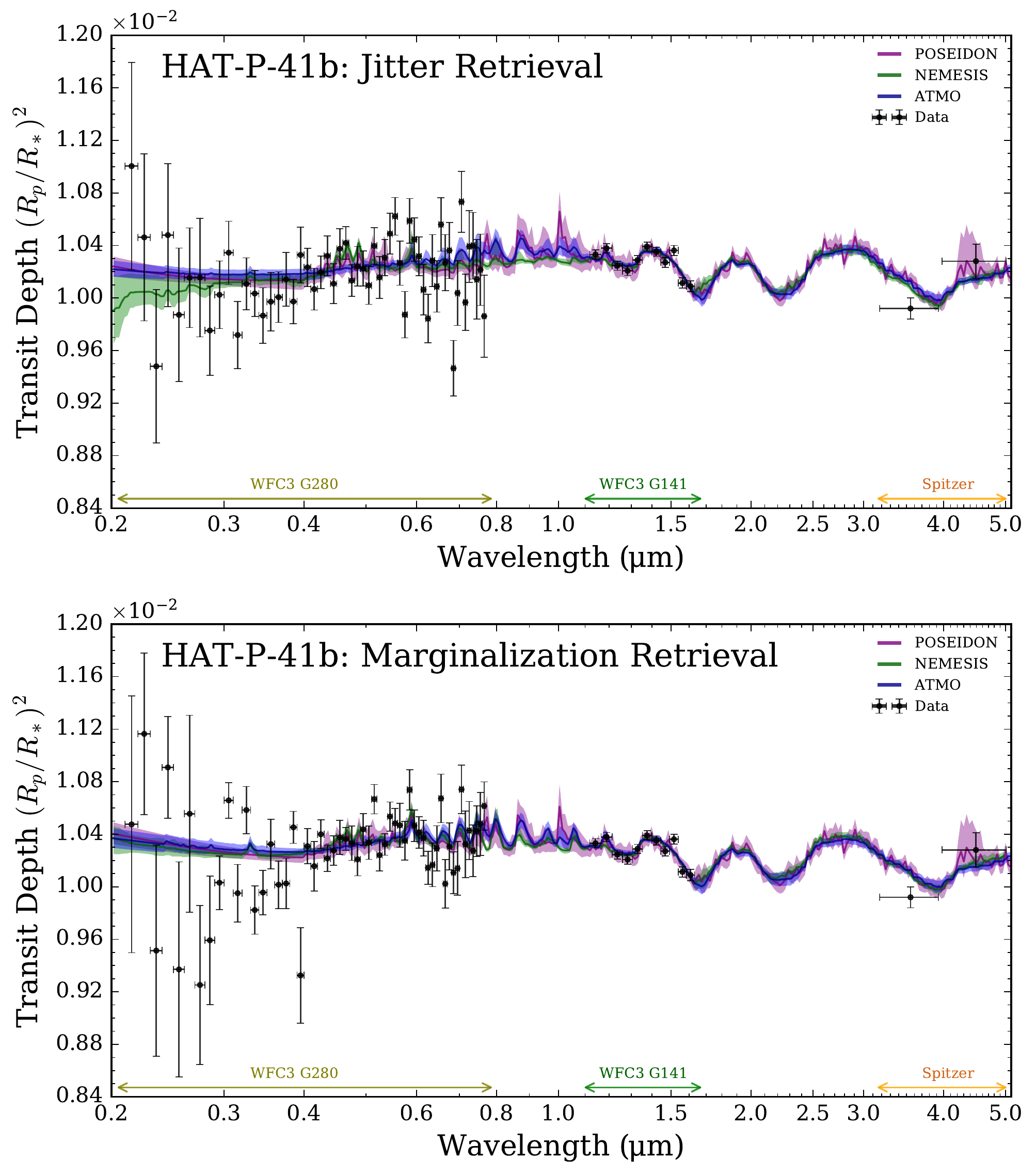}
\caption{Atmospheric retrievals of HAT-P-41b's transmission spectrum. Top: jitter decorrelation reduction. Bottom: systematic marginalization reduction. Each panel shows the median retrieved spectrum (solid lines) and 1$\sigma$ confidence regions (shading) from three retrieval codes: POSEIDON (purple), NEMESIS (green), and ATMO (blue) \citep{macdonald2017,barstow2017,Wakeford2017Science}. All models are binned to a common spectral resolution ($R = 100$) for clarity. The three codes achieve an excellent fit across the full wavelength range, concurring on the presence of at least one significant opacity source in the visible and near-UV (H$^{-}$, Na, CrH, VO, or AlO - see Appendix~\ref{appendix_B} for a spectral decomposition), H$_2$O in the near-infrared, and an absence of clouds in the observable atmosphere. This interpretation holds for both reductions.} \label{fig:retrievals}
\end{figure*}

Atmospheric retrievals relax many aforementioned assumptions, such as chemical equilibrium, opting instead to parametrize the atmospheric state. Bayesian sampling techniques explore millions of potential states, comparing their resultant transmission spectra with observations to derive posterior probability distributions for each model parameter. This inverse approach allows atmospheric properties, such as chemical abundances, temperature profiles, and cloud properties, to be retrieved directly from the data.

Our retrieval philosophy employs two central principles to ensure robust atmospheric inferences. First, we conduct retrievals for each data reduction separately, establishing the sensitivity of derived atmospheric properties to different reduction techniques. Secondly, three different retrieval codes are independently applied to each data reduction: POSEIDON \citep{macdonald2017}, NEMESIS \citep{irwin08,barstow2017,Krissansen-Totton2018}, and ATMO \citep{amundsen2014,tremblin2015,Wakeford2017Science}. We can thereby quantify the impact on our results of differing modelling choices (e.g. molecular line lists, temperature profiles, cloud parametrizations).

We consider an extensive range of potential atmospheric components. Our investigations include opacity due to the following chemical species: H$_2$, He, H$^{-}$, Na, K, Li, TiO, VO, AlO, SiO, TiH, CrH, FeH, AlH, CaH, SiH, H$_2$O, CH$_4$, CO, CO$_2$, NH$_3$, HCN, NO, H$_2$S, SH, PH$_3$, and C$_2$H$_2$. Isothermal (NEMESIS and ATMO) and non-isothermal temperature structures (POSEIDON, \citealt{madhusudhan2009}) were considered. Three cloud models were examined: (i) an opaque cloud deck with a vertically-uniform haze (ATMO, \citealt{Wakeford2017Science}); (ii) a single cloud deck, with variable top and base pressures, and power law extinction with a variable index (NEMESIS, \citealt{barstow2017}); and (iii) patchy clouds and haze around the terminator (POSEIDON, \citealt{macdonald2017}). Although these cloud prescriptions do not explicitly include Mie scattering calculations, our model fitting exercise with CARMA (see Section \ref{subsec:cloud})
demonstrates that more complex microphysical prescriptions for clouds do not fully explain the observed spectrum of HAT-P-41b. The influence of stellar contamination was also considered in the retrieval process (NEMESIS, e.g. \citealt{pinhas2018}).
Iterative expansion of the considered molecules was performed between the three retrieval codes until a minimal basis set was identified. All three codes use nested sampling to explore the parameter space, either via MultiNest (NEMESIS and POSEIDON, \citealt{feroz2008,feroz2009,feroz2013,buchner2014}) or dynesty (ATMO, \citealt{speagle2020}).

Our retrieved transmission spectra are compared to the observations of HAT-P-41b in Figure~\ref{fig:retrievals}. Contrasting with the forward models of the previous sections, the retrievals prefer a combination of gas phase optical opacity sources instead of clouds. Specifically, at least one of H$^{-}$, AlO, CrH, or VO\footnote{Absorption data for H$^{-}$, AlO, CrH and VO are taken from \cite{john88}, \cite{jt598}, \cite{MOLLIST}, and \cite{jt644}, respectively.}, in addition to Na, are required to explain the WFC3 UVIS G280 observations. By invoking chemical species with strong near-UV to visible opacities, but weak infrared opacities, the infrared observations can be well fit by H$_2$O alone. In particular, there is a clear preference for the bound-free opacity of the hydrogen anion, H$^{-}$, which provides a smooth continuum across the UVIS range before falling off rapidly as the ionization threshold ($\sim$ 1.64~$\micron$) is approached. This continuum has similar spectral characteristics to a cloud deck across the UVIS range, potentially explaining the preference for clouds in our forward models (which do not include H$^{-}$). The other inferred opacity sources are somewhat sensitive to the retrieval code\footnote{Note that NEMESIS does not currently support CrH, while ATMO does not support AlO.} and chosen data reduction, as we demonstrate below. Nevertheless, all three retrieval codes agree on the interpretation of a strong near-UV to visible chemical opacity source, without the need for clouds in the observable atmosphere.

\subsection{The Atmospheric Composition of HAT-P-41b \label{subsec:composition}}

We detect the presence of H$_2$O in HAT-P-41b's atmosphere at $\gtrsim 5\sigma$ confidence (jitter: $6.2\sigma$; marginalization: $5.1\sigma$). The visible and near-UV observations additionally require at least one other prominent opacity source, with candidates identified as H$^{-}$, CrH, AlO, VO, or Na. Besides these species, our initial retrievals - ranging in complexity from 12 to 37 free parameters - included many parameters left largely unconstrained by the present observations (e.g. temperature structure and cloud properties). Consequently, the best-fitting $\chi^{2}_{\nu}$ ranged from $\sim$ 1.7 to 3.0 (see Appendix~\ref{appendix_A}). We therefore constructed an 8-parameter `minimal' model, including only those parameters found necessary (via the Bayesian evidence) to explain the observations: an isothermal, clear, atmosphere with H$_2$O, H$^{-}$, Na, CrH, AlO, and VO. This best-fitting model attains $\chi^{2}_{\nu} =$ 1.50 and 1.72 for the jitter and marginalization cases, respectively.

The $\chi^{2}_{\nu}$ values obtained by our best-fitting retrieval models demonstrate a greatly improved quality of fit compared to the chemical equilibrium and self-consistent cloud models considered in Sections~\ref{subsec:forward} and \ref{subsec:cloud}. However, we note that our chi-squared values still suggest some level of tension between the data and models. Under frequentist metrics one could still consider rejecting all the models presented here, but this argument assumes that both the data and models perfectly capture all noise sources and atmospheric physics. Tactics such as error bar inflation \citep[as done in studies such as][]{line2015, zhang2018, colon2020} and incorporating models of additional complexity \citep[e.g. see discussion in][]{gibson2014, parviainen2018} could be used to further reduce these chi-squared values, but could obfuscate our physical interpretation for HAT-P-41b from these data and associated inter-model comparisons. It is important to note that both H$_2$O and UV-optical opacity sources are independently required to explain the observations, irrespective of chi-squared tests, as established by our Bayesian model comparisons.

With reference to our best-fitting `minimal' model, we conducted a series of Bayesian model comparisons with POSEIDON to compute detection significances for each inferred UV-optical chemical species. The jitter reduction yields moderate evidence for H$^{-}$ ($2.9\sigma$), weak evidence for CrH ($2.7\sigma$) and AlO ($2.4\sigma$), and a tentative hint of Na ($2.0\sigma$). The marginalization reduction yields moderate evidence for Na ($2.9\sigma$), weak evidence for H$^{-}$ ($2.6\sigma$) and CrH ($2.5\sigma$), and tentative hints of AlO ($1.9\sigma$) and VO ($1.7\sigma$). The specific spectral features giving rise to these inferences are shown in Appendix~\ref{appendix_B} (see Figure~\ref{fig:opacty_contributions}). The differing significances for Na and VO highlight the sensitivity of some atmospheric inferences to specific data reductions. However, our most rigorous conclusion holds for both reductions: at least one of H$^{-}$, CrH, AlO, or VO is required at $> 5\sigma$ (jitter: $5.1\sigma$; marginalization: $5.5\sigma$) to explain HAT-P-41b's transmission spectrum.

\begin{figure*}[ht!]
\includegraphics[width=\textwidth]{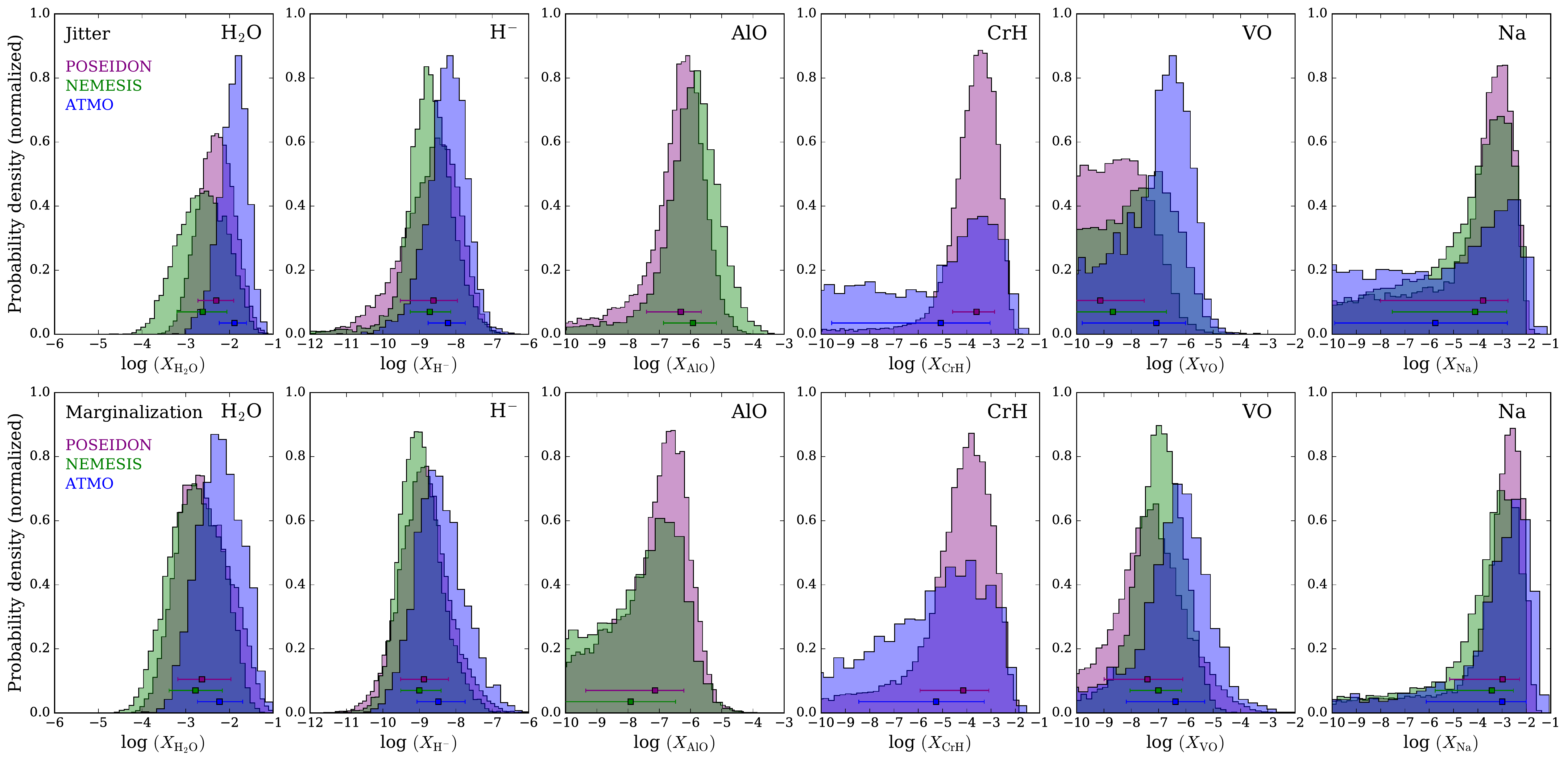}
\caption{Retrieved atmospheric composition of HAT-P-41b. The histograms show marginalized posterior probability distributions for the volume mixing ratios of each chemical species inferred by at least one retrieval code. The posteriors from POSEIDON (purple), NEMESIS (green), and ATMO (blue) are compared. Where a retrieval code does not include a given species, no histogram is shown. The error bars give the median retrieved abundances and $\pm 1\sigma$ confidence levels. The retrievals agree on a $5\sigma$ detection of H$_2$O and evidence of at least one visible to near-UV absorber at $3\sigma$ confidence. The retrieved abundances from each code, and for each data reduction, are broadly consistent within their respective $1\sigma$ confidence regions.} \label{fig:abundances}
\end{figure*}

The retrieved abundances for each inferred chemical species\footnote{The abundances for other included chemical species are constrained only by upper bounds - see the \href{https://doi.org/10.5281/zenodo.4023155}{supplementary material}.} are shown in Figure~\ref{fig:abundances}. All three codes provide precise H$_2$O abundances. Across both data reductions and all three retrieval codes, the retrieved H$_2$O abundance spans $\log(\rm{H_2 O}) \approx$ -3.4 to -1.6 (with a mean precision of 0.5 dex). The H$_2$O abundances from ATMO are $\gtrsim$ 0.4 dex higher than those retrieved by NEMESIS and POSEIDON. Nevertheless, the retrieved abundances are consistent to $1\sigma$ - see Table~\ref{tab:retrieval_results}. These results illustrate that the chosen data reduction technique at visible wavelengths can alter retrieved H$_2$O abundances by $\lesssim$ 0.5 dex. We additionally report a precise H$^{-}$ abundance, consistent across all three retrieval codes and both data reductions, with a mean value of $\log(\rm{H^{-}}) = -8.65 \pm 0.62$. Abundance constraints for the other inferred species are relatively weak, and can vary significantly between different retrieval codes and data reductions (e.g. the VO posteriors in Figure~\ref{fig:abundances}). However, the derived H$_2$O and H$^{-}$ abundances are robust to both modelling choices and data reduction techniques. We note that these abundances are with respect to the original `full' retrievals (see Table~\ref{tab:retrieval_results}), ensuring that model uncertainty (overlapping absorption features, cloud-chemistry degeneracies, etc.) are automatically encoded in the quoted values. However, biases may still arise from neglected model complexity. In particular, retrieved H$_2$O and H${-}$ abundances from 1D retrieval methods can be biased by up to 1 dex if compositional gradients arise between the morning and evening terminators \citep{MacDonald2020}. Such a bias would cause a slight overestimation in our retrieved H$_2$O abundances, and underestimate in our H$^{-}$ abundances \citep[see][Figure 3]{MacDonald2020}.

The precise H$_2$O abundances we obtain can be converted into estimates of the atmospheric metallicity. By `metallicity', we refer to the atmospheric O/H ratio relative to that of its star ([Fe/H] = 0.21, \citet{stassun2017}). The retrieved molecular abundances are mapped into O/H ratios as in \citet{macdonald2019}. We note that our metallicities should only be considered accurate to a factor of two\footnote{At HAT-P-41b's temperature, approximately half of the atmospheric O is expected to reside in CO \citep{madhusudhan2012}.}, given that the current data are insensitive to CO. The metallicities derived by POSEIDON and NEMESIS are consistent with the stellar metallicity of HAT-P-41 for the marginalization reduction: $1.73^{+6.26}_{-1.24} \times$ stellar and $1.22^{+3.80}_{-0.92} \times$ stellar, respectively. The ATMO retrievals find $\sim 4 \times$ higher metallicities (due to the aforementioned higher H$_2$O abundances): $4.35^{+10.43}_{-2.99} \times$ stellar for the marginalization reduction. Comparatively, the jitter reduction favours slightly super-stellar metallicities: $3.65^{+5.59}_{-2.25} \times$ stellar (POSEIDON), $1.79^{+4.66}_{-1.33} \times$ stellar (NEMESIS), and $9.57^{+8.43}_{-5.28} \times$ stellar (ATMO). Nevertheless, all derived metallicities are consistent with the stellar value to $2\sigma$. Overall, the metallicities derived by the different retrieval codes are consistent with each other. We conclude that the metallicity of HAT-P-41b's atmosphere is consistent with being stellar, or slightly super-stellar, in line with expected mass-metallicity trends. 

\subsection{The Case for Disequilibrium Chemistry}\label{sec:disequil}

Our retrieved abundances for visible to near-UV absorbers require an atmosphere in chemical disequilibrium. Here we consider disequilibrium mechanisms that might enhance the atmospheric abundance of AlO, CrH, and H$^{-}$ in HAT-P-41b's atmosphere, the species common to both the jitter and marginalization data reduction analyses (see discussion in section \ref{subsec:composition}). At the retrieved terminator temperature\footnote{Our retrieved temperature profiles are available in the \href{https://doi.org/10.5281/zenodo.4023155}{supplementary material}.} ($\sim 1000 \pm 200$~K), the equilibrium abundances of CrH and AlO for a solar metallicity atmosphere are respectively $\sim$ 7 and 5 orders of magnitude below those we infer \citep{woitke2018}, which represents a significant discrepancy between the retrieved and equilibrium chemistry predicted abundances as well as the atmospheric metallicity we infer from the retrieved H$_2$O abundance. Even at depth in HAT-P-41b's atmosphere ($P\sim1$~bar, $T\sim$~2000~K) the expected abundances for CrH and AlO are several orders of magnitude less than the retrieved abundances \citep{woitke2018}, which makes it difficult for a mechanism such as vertical quenching to account for their disequilibrium abundances. Given this difficulty in identifying a plausible mechanism for a significant enhancement of CrH or AlO in HAT-P-41b's atmosphere we turn our attention to H$^{-}$, which had the highest statistical significance (2.6--2.9$\sigma$) of the UV-optical absorbing species we considered in our retrievals. Under equilibrium thermochemistry, H$^{-}$ is not expected to exist in appreciable ($\geq$10$^{-9}$ mole fraction) quantities for $T \lesssim 2500$~K \citep{kitzmann2018}. We verified the need for H$^{-}$ opacity by conducting an additional ATMO retrieval with chemical equilibrium enforced. This retrieval raised the atmospheric temperature to $\gtrsim 2500$~K, far above $T_{\rm{eq}}$ and predictions from the GCM (Figure~\ref{fig:gcm_profiles}), to create H$^{-}$ opacity with an abundance consistent with that found by our free retrievals (see the \href{https://doi.org/10.5281/zenodo.4023155}{supplementary material}).

We explored the potential for photochemistry to enhance the abundance of H$^{-}$ in a hot Jupiter with $T \lesssim 2500$~K like HAT-P-41b. 
In the pressure regions probed by transits ($\sim$1~mbar), \citet{Lavvas2014} show that electron densities will be roughly an order of magnitude higher than predictions from thermal ionization only. It is also expected that photochemistry will enhance the neutral H abundance by many orders of magnitude higher than those predicted by equilibrium chemistry \citep[e.g.][]{liang2003, moses2011}. This expected increase in electron densities and neutral H abundance in the atmosphere of a hot Jupiter like HAT-P-41b increases the chance for production of H$^{-}$ via radiative electron attachment ($\mathrm{H+e^-\rightarrow H^- + \gamma}$), collisional electron attachment ($\mathrm{H+e^-+M\rightarrow H^-+M}$), and dissociative electron attachment ($\mathrm{H_2+e^-\rightarrow H +H^-}$). For conditions relevant to HAT-P-41b, the latter process likely dominates, due to the prevalence of H$_2$ and the moderately large rate coefficient of order of 10$^{-13}$~cm$^{3}$~s$^{-1}$ \citep{Janev2003}. The destruction of H$^{-}$ can occur through collisional detachment (e.g. $\mathrm{H^{-}+H \rightarrow H_2 + e^{-}}$ or $\mathrm{H^{-}+H_2O \rightarrow OH^{-} + H_2}$) 
with rate coefficients for such reactions on the order of a few $\times$  10$^{-9}$~cm$^{3}$~s$^{-1}$ \citep{bruhns2010, martinez2010}. From the 0.1 mbar atmospheric density from the global-average GCM results described above and the mixing ratios of e$^{-}$, H, and H$_2$ from the model of \citet{Lavvas2014}, we can estimate number densities for e$^{-}$, H, and H$_2$ near the $\sim$0.1~mbar level of HAT-P-41b's atmosphere to be $\sim$10$^9$, 2 $\times$ 10$^{13}$, and 6 $\times$ 10$^{14}$~cm$^{-3}$ respectively. Assuming that in steady state the H$^{-}$ production and loss rates balance each other, and considering production solely by H$_2$ dissociative electron attachment and destruction solely by collisional detachment with atomic H, we estimate a number density for H$^{-}$ on the order of 10$^6$~cm$^{-3}$. This value corresponds to a H$^{-}$ mixing ratio of 2 $\times$ 10$^{-9}$, consistent with our retrieved value for the abundance of H$^{-}$ and roughly six orders of magnitude larger than expectations from equilibrium chemistry for a planet like HAT-P-41b. This order-of-magnitude estimate highlights that significant enhancement of H$^{-}$ due to photochemical processes is likely present in HAT-P-41b and many other exoplanet atmospheres, thus shaping their UV-optical spectra. The production of H$^{-}$ will be particularly enhanced for hot planets that receive a high extreme UV flux from their host stars and have Na in the gas phase, which increases electron production \citep{Lavvas2014}. Such conditions are expected for HAT-P-41b \citep[this work;][]{hartman2012,linsky2014} and other hot Jupiters orbiting F-stars. 

\section{Discussion and Conclusions}

Our analysis of the transmission spectrum of the hot Jupiter HAT-P-41b from 0.2-5.0~$\mu$m represents one of the most comprehensive explorations of an exoplanet atmosphere to date. In particular, our analysis includes new high-precision information at UV/NUV wavelengths provided by Hubble's WFC3 UVIS G280 grism. We leveraged multiple reductions of the WFC3 UVIS G280 data and multiple spectral analysis tools to obtain a more complete and robust picture of the physical and chemical processes at work in HAT-P-41b's atmosphere. We find that:

\begin{itemize}
    \item The presence of a significant cloud deck, composed of aluminum bearing species, provides a plausible explanation for the UV, optical and 4.5~$\mu$m portions of HAT-P-41b's transmission spectrum, but is discrepant with observations in the NIR (1.1-1.7~$\micron$) and at 3.6~$\micron$. This highlights the need for broad wavelength coverage from the UV to IR to constrain atmospheric properties, in particular aerosols, in exoplanet atmospheres.
    \item Our use of multiple reductions of the {\it Hubble} WFC3 UVIS G280 observations and multiple interpretation methods shows that in most areas we obtain a consistent picture for HAT-P-41b's atmosphere, which proves the robustness of our results. We also highlight that potentially spurious conclusions can be drawn when relying on single data reduction and interpretation techniques. In particular, the presence of VO in HAT-P-41b's atmosphere is more strongly preferred for data where the marginalization approach is used to correct for systematics in the WFC3 UVIS G280 data and in retrievals performed with the ATMO model. 
    \item We find evidence for the presence of the hydrogen anion, H$^{-}$, in HAT-P-41b and provide precise constraints for its abundance: $\log(\rm{H^{-}}) = -8.65 \pm 0.62$. This represents an abundance for H$^{-}$ several orders of magnitude larger than what would be expected via equilibrium chemistry for HAT-P-41b given its equilibrium temperature of $\sim$1700-1950~K, retrieved temperature of $\sim$1000~K, and predictions of its thermal structure from one and three-dimensional models. This points to the possibility of a not yet considered disequilibrium chemistry process in hot Jupiter atmospheres, which may be driven by intense UV radiation from stars like HAT-P-41b's F-type host star. Order-of-magnitude calculations from section \ref{sec:disequil} demonstrate that photochemical and collisional processes on hot hydrogen-dominated exoplanets can readily supply the necessary amount of H$^{-}$.
\end{itemize}

In the future, the James Webb Space Telescope (JWST) will provide the exoplanet community with high-precision spectroscopic observations of exoplanet atmospheres spanning 0.6-14~$\mu$m \citep{beichman2014}. The complexities encountered in the reduction, analysis, and interpretation of the HAT-P-41b observations presented here will also be encountered with JWST observations; as such this study serves both to highlight the challenges and provides a needed test-bed for future transiting exoplanet observations. Additionally as highlighted by this work, observations in the UV/NUV critically complement atmospheric transmission observations in the optical and infrared, probing the presence of a range of chemical species and giving insights into processes occurring in upper atmospheres such as stratospheric heating and photochemistry.  

\acknowledgments{
This research is based on observations made with the NASA/ESA Hubble Space Telescope obtained from the Space Telescope Science Institute, which is operated by the Association of Universities for Research in Astronomy, Inc., under NASA contract NAS 5–26555. These observations are associated with programs GO-15288 and GO-14767. This work is based in part on observations made with the Spitzer Space Telescope, which is operated by the Jet Propulsion Laboratory, California Institute of Technology under a contract with NASA. These observations are associated with program 13044. We thank Patrick Irwin for the use of NEMESIS, and Jake Taylor for assistance with the inclusion of H$^{-}$ opacity within the NEMESIS forward model.

\software{IDL Astronomy user's library \citep{IDL_ref}, NumPy \citep{numpy}, SciPy \citep{scipy}, MatPlotLib \citep{matplotlib}, AstroPy \citep{astropy}, Photutils \citep{photutils}, MultiNest \citep{feroz2008,feroz2009,feroz2013}, PyMultiNest \citep{buchner2014}, dynesty \citep{speagle2020}, ATMO \citep{amundsen2014,tremblin2015,tremblin2016,Wakeford2017Science,goyal2018}, NEMESIS \citep{irwin08, barstow2017}, POSEIDON \citep{macdonald2017}.}
\vspace{5mm}
\facilities{HST(WFC3), Spitzer(IRAC)}}



\appendix
\section{WFC3-IR G141 transmission comparison}\label{appendix_0}
\begin{figure*}[h!]
\includegraphics[width=\textwidth]{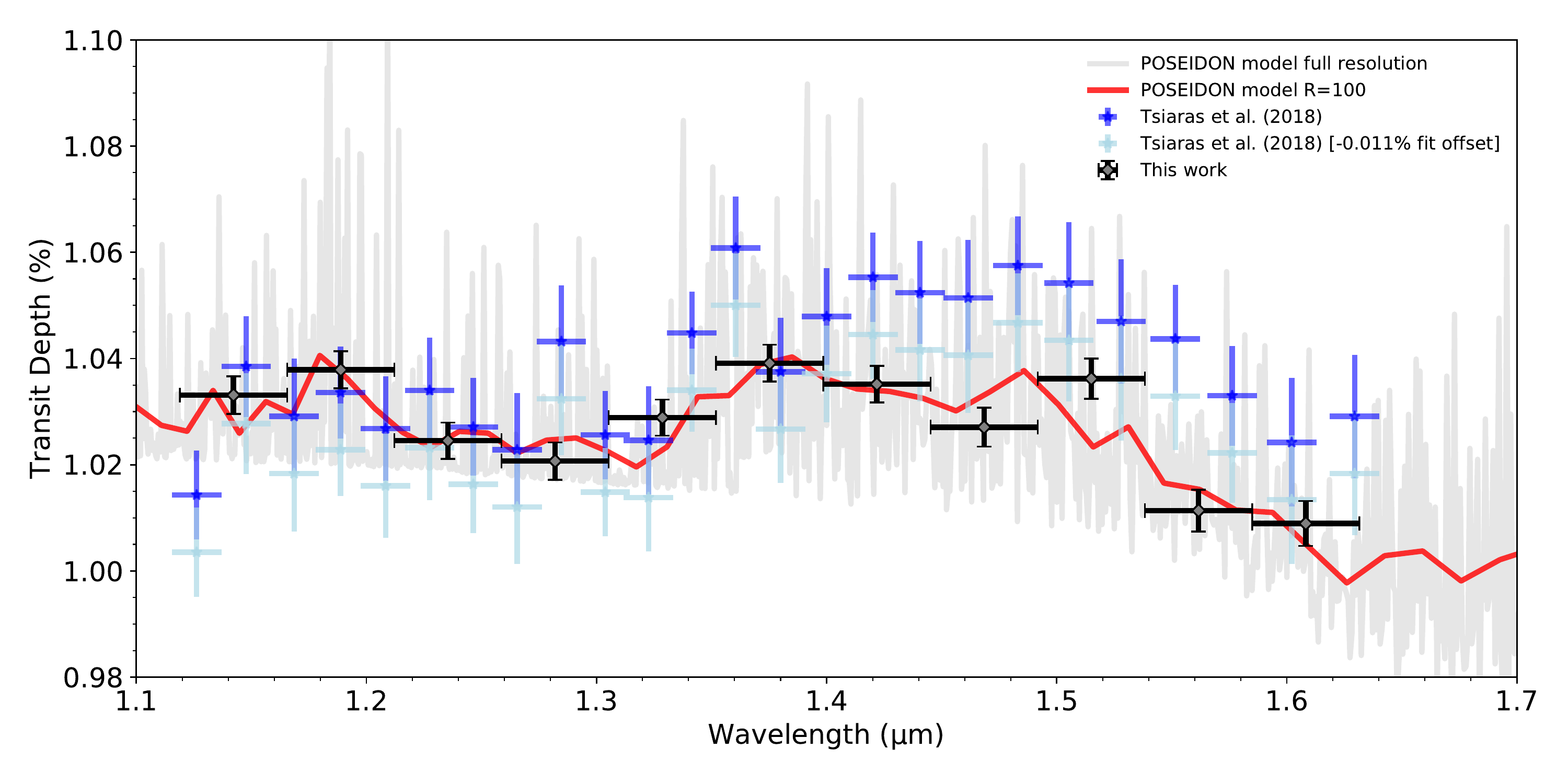}
\caption{Measured near-IR transmission spectrum of HAT-P-41b from HST/WFC3-IR G141 from this work (black) and the best fitting POSEIDON model using the jitter data (corresponding to the 6 parameter `minimal' model) (grey \& red). Also plotted are the published spectra by \citet{Tsiaras2018}[T18] (dark blue) and T18 shifted in altitude by -0.011\% to the model by minimizing the chi-squared (light blue). This demonstrates the differences and similarities to the previously published spectrum and the reduction presented in this paper for use in the full UV-IR transmission spectrum.
} \label{fig:tsiaras_comparison}
\end{figure*}

Figure \ref{fig:tsiaras_comparison} shows the comparison between the reduction presented here and by \citet{Tsiaras2018}[T18] for {\it Hubble's} WFC3 G141 grism observations of HAT-P-41b. To more accurately compare the shape of the transmission spectra the T18 spectrum was shifted down in altitude by 0.011\%. This offset is likely caused by differences in the system parameters used in the light curve fitting stage of the analysis between T18 and this study which leverages updated parameters for the HAT-P-41b system published in \citet{wakeford2020}.
Using common system parameters across all datasets considered in this study ensures a consistent analysis across the entire transmission spectrum and therefore in the interpretation of the planetary atmosphere. In addition to an offset there is a small difference in the slope of the two transmission spectra that is most likely caused by differences in the assumed stellar properties and limb-darkening model employed. Again for consistency, we utilize the same models, methods, and system parameters (including updated stellar properties based on \citet{morrell2019}) presented in \citet{wakeford2020} to account for limb-darkening in the WFC3 G141 grism spectrum of HAT-P-41b.

In Fig. \ref{fig:tsiaras_comparison} we compare the transmission spectra to the POSEIDON model that was fit to the jitter decorrelated transmission spectrum including the near-IR data. For the near-IR data only, when compared to the POSEIDON model binned to the resolution of the data, the data from this work has a $\chi^2_\nu$ = 1.38 with 11 degrees of freedom (DOF), the published spectra from T18 without an offset has $\chi^2_\nu$  = 2.70 with 25 DOF, and when shifted in altitude to the model $\chi^2_\nu$  = 1.53 with 24 DOF. This demonstrates the similarity in the shape of the reduced transmission spectra and highlights the effect of offsets between different analysis techniques.

\section{Full atmospheric retrieval results and model comparison} \label{appendix_A}

Table~\ref{tab:retrieval_results} summarises the retrieved values for the 8 common parameters our atmospheric retrievals found necessary to explain HAT-P-41b's observed transmission spectrum. All three retrieval codes reach good agreement, despite their varying complexity, with all obtaining a precise H$^{-}$ abundance constraint and inferring a stellar (or slightly super-stellar) O/H ratio. The Bayesian evidence and reduced chi-squared statistics both prefer a `minimal' model (with only the 8 parameters in Table~\ref{tab:retrieval_results}), which nevertheless yields consistent parameter constraints with the more complex retrieval models.

\section{Spectral evidence of UV-visible absorbers in HAT-P-41b's atmosphere} \label{appendix_B}

\begin{figure*}[ht!]
\includegraphics[width=\textwidth]{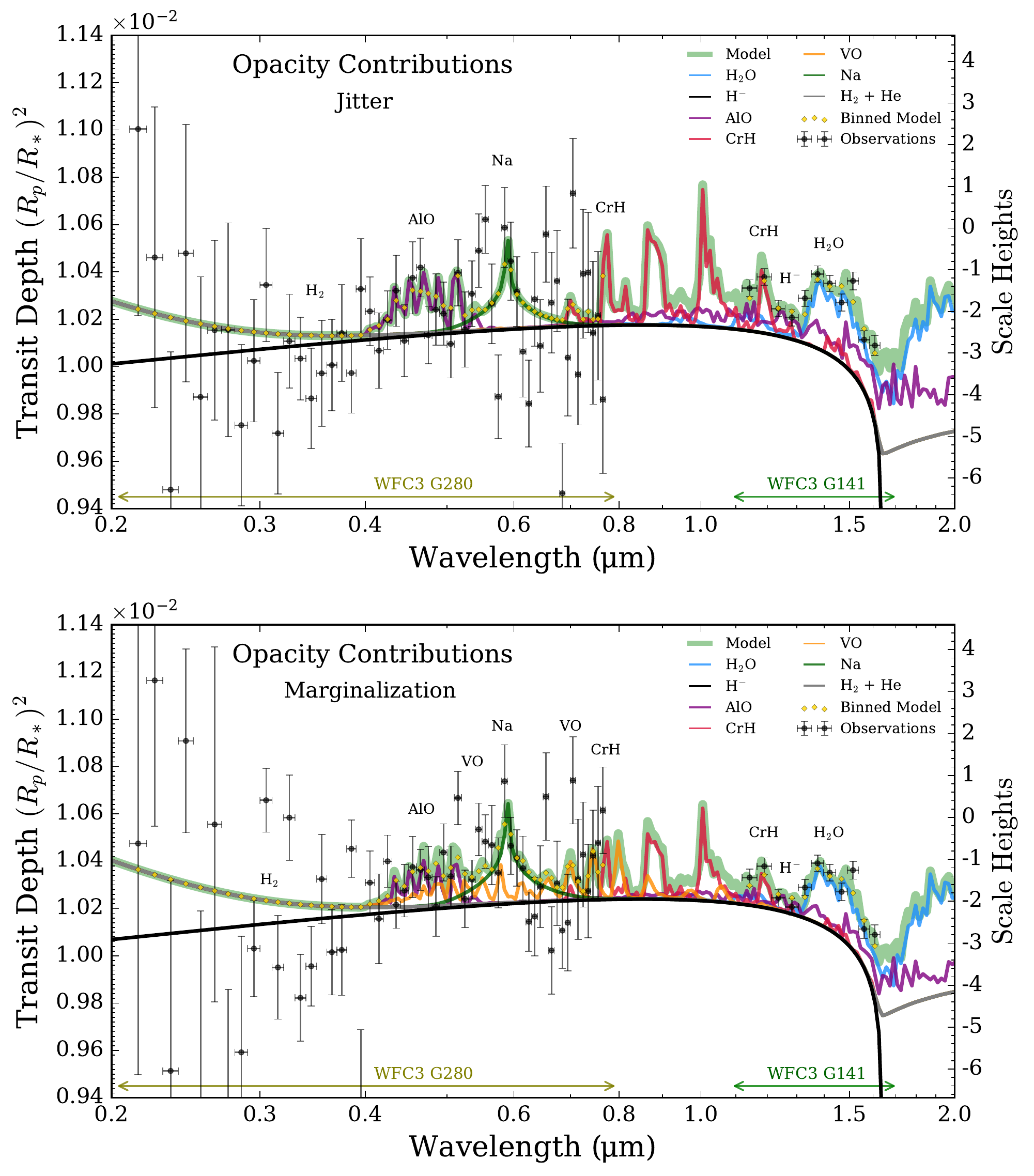}
\caption{Opacity contributions to the best-fitting model transmission spectra of HAT-P-41b. Each panel shows the maximum likelihood spectrum (green shading) from the `minimal' POSEIDON retrievals for each WFC3 G280 data reduction (top: jitter decorrelation; bottom: systematic marginalization). The UV-visible spectrum is shaped by H$^{-}$ bound-free opacity (black curve) from $\sim$ 0.4-1.6~$\micron$. The opacity contributions of other retrieved species (H$_2$O, Na, CrH, AlO, and VO) are depicted relative to the H$^{-}$ continuum (colored curves). H$_2$ Rayleigh scattering contributes opacity for wavelengths $\lesssim$ 0.4~$\micron$. The best-fitting model, binned to the resolution of each set of observations, is overlaid for comparison (gold diamonds).} \label{fig:opacty_contributions}
\end{figure*}

Figure~\ref{fig:opacty_contributions} shows a spectral decomposition of our best-fitting model transmission spectra. This illustrates which features in the observations are attributed to specific chemical species. Note that these opacity contributions are \emph{relative} to the H$^{-}$ continuum, which serves to `boost' the transit depth contributions of other absorbing species. Both data reductions produce similar best-fitting models, with the only difference being a slight preference to include VO for the systematic marginalization reduction.

\begin{deluxetable*}{lcccccccccc}[ht!] \label{tab:retrieval_results}
    \tablecaption{Atmospheric Retrieval Analysis Summary}
    \tablewidth{0pt}
    \tablehead{
    \multicolumn{1}{c}{\hspace{-2.1em} \bfseries Data Reduction} & \phm{-} &
    \multicolumn{4}{c}{\bfseries Jitter} & \phm{-} & \multicolumn{4}{c}{\bfseries Marginalization}\\ \cmidrule{1-1} \cmidrule{3-6} \cmidrule{8-11}
    Retrieval & & POSEIDON & NEMESIS & ATMO & `Minimal' & \phm{-} & POSEIDON & NEMESIS & ATMO & `Minimal'
    }
    \startdata \\[-8pt]
    \textbf{Parameters} \\ 
    \hspace{0.5em} $T_{\rm{1 \, mbar}}$ (K) & & $1148^{+194}_{-182}$ & $1005^{+193}_{-178}$ & $1088^{+196}_{-153}$ & $1149^{+248}_{-209}$ & & $988^{+209}_{-192}$ & $938^{+194}_{-161}$ & $936^{+272}_{-233}$ & $1124^{+274}_{-250}$ \\
    \hspace{0.5em} $R_{\rm{p, \, ref}}$ ($R_J$) & & $1.58^{+0.02}_{-0.02}$ & $1.59^{+0.01}_{-0.02}$ & $1.64^{+0.01}_{-0.01}$ & $1.58^{+0.02}_{-0.02}$ & & $1.60^{+0.01}_{-0.02}$ & $1.59^{+0.01}_{-0.01}$ & $1.65^{+0.01}_{-0.01}$ & $1.59^{+0.02}_{-0.02}$ \\
    \hspace{0.5em} log($X_{\rm{H_2 O}}$) & & $-2.31^{+0.41}_{-0.42}$ & $-2.61^{+0.55}_{-0.58}$ & $-1.89^{+0.27}_{-0.35}$ & $-2.57^{+0.49}_{-0.47}$ & & $-2.63^{+0.67}_{-0.55}$ & $-2.78^{+0.61}_{-0.60}$ & $-2.23^{+0.53}_{-0.50}$ & $-3.38^{+0.62}_{-0.55}$ \\
    \hspace{0.5em} log($X_{\rm{H^{-}}}$) & & $-8.62^{+0.66}_{-0.91}$ & $-8.72^{+0.58}_{-0.53}$ & $-8.22^{+0.48}_{-0.54}$ & $-8.64^{+0.57}_{-0.55}$ & & $-8.88^{+0.68}_{-0.64}$ & $-9.00^{+0.60}_{-0.51}$ & $-8.48^{+0.74}_{-0.59}$ & $-9.26^{+0.56}_{-0.55}$ \\
    \hspace{0.5em} log($X_{\rm{AlO}}$) & & $-6.31^{+0.66}_{-1.10}$ & $-5.92^{+0.75}_{-0.94}$ & --- & $-6.53^{+0.73}_{-1.44}$ & & $-7.13^{+0.92}_{-2.22}$ & $-7.92^{+1.44}_{-2.88}$ & --- & $-7.24^{+0.78}_{-1.59}$ \\
    \hspace{0.5em} log($X_{\rm{CrH}}$) & & $-3.60^{+0.74}_{-0.99}$ & --- & $-5.08^{+2.03}_{-4.49}$ & $-4.08^{+0.93}_{-1.31}$ & & $-4.15^{+1.05}_{-1.78}$ & --- & $-5.27^{+1.99}_{-3.17}$ & $-4.94^{+1.27}_{-1.53}$ \\
    \hspace{0.5em} log($X_{\rm{VO}}$) & & $-9.13^{+1.60}_{-1.71}$ & $-8.67^{+1.96}_{-2.69}$ & $-7.08^{+1.07}_{-2.72}$ & $-9.67^{+1.60}_{-1.57}$ & & $-7.40^{+1.29}_{-1.59}$ & $-7.01^{+0.85}_{-1.03}$ & $-6.38^{+1.07}_{-1.79}$ & $-8.23^{+1.06}_{-1.82}$ \\
    \hspace{0.5em} log($X_{\rm{Na}}$) & & $-3.78^{+1.02}_{-4.22}$ & $-4.11^{+1.30}_{-3.40}$ & $-5.75^{+2.96}_{-4.15}$ & $-4.45^{+1.27}_{-4.52}$ & & $-2.98^{+0.70}_{-2.18}$ & $-3.42^{+0.89}_{-2.33}$ & $-3.00^{+0.99}_{-3.12}$ & $-3.42^{+0.89}_{-1.08}$ \\[3pt]
    \midrule
    \textbf{Derived Properties} \\
    \hspace{0.5em} O/H ($\times$ stellar) & & $3.65^{+5.59}_{-2.25}$ & $1.79^{+4.66}_{-1.33}$ & $9.57^{+8.43}_{-5.28}$ & $1.96^{+4.07}_{-1.30}$ & & $1.73^{+6.26}_{-1.24}$ & $1.22^{+3.80}_{-0.92}$ & $4.35^{+10.43}_{-2.99}$ & $0.31^{+0.96}_{-0.22}$ \\[3pt]
    \midrule
    \textbf{Statistics} \\ 
    \hspace{0.5em} ln(Evidence) & & $473.9$ & $159.8$ & $473.3$ & $478.9$ & & $472.6$ & $152.5$ & $473.9$ & $477.8$ \\
    \hspace{0.5em} $\chi^2_{\nu, \, \rm{min}}$ & & $2.55$ & $1.90$ & $1.73$ & $1.50$ & & $3.02$ & $2.37$ & $1.97$ & $1.72$ \\
    \hspace{0.5em} $N_{\rm param}$ & & $37$ & $17$ & $12$ & $8$ & & $37$ & $17$ & $12$ & $8$ \\
    \hspace{0.5em} d.o.f. & & $32$ & $52$ & $57$ & $61$ & & $32$ & $52$ & $57$ & $61$ \\[3pt]
    \enddata 
    \tablecomments{All retrievals here have `free composition', without the assumption of chemical equilibrium. The `minimal' model contains only the 8 free parameters found necessary to fit either data reduction (i.e. those listed in the table). Only parameters with bounded constraints (i.e. both lower and upper bounds) are included - see the online \href{https://doi.org/10.5281/zenodo.4023155}{supplementary material} for full posterior distributions.  $R_{\rm{p, \, ref}}$ is defined at $P =$ 10~bar for NEMESIS and POSEIDON, and 1~mbar for ATMO. The NEMESIS retrievals use a different evidence normalizing factor to ATMO and POSEIDON. The stellar O/H is assumed equal to HAT-P-41's stellar [Fe/H] (0.21, \citet{stassun2017}).  Equilibrium retrievals with similar complexity to the minimal model are omitted, due to their relatively poor fits (e.g. an ATMO equilibrium retrieval for the marginalization reduction obtained $\chi^2_{\nu, \, \rm{min}} = 2.38$ for 62 degrees of freedom). }
    \vspace{-10pt}
\end{deluxetable*}

\bibliography{hat41_theory}{}
\bibliographystyle{aasjournal}

\end{document}